# Primer - A Tool for Testing Honeypot Measures of Effectiveness

Jason M. Pittman, Kyle Hoffpauir, and Nathan Markle
High Point University, High Point, NC
Email: jpittman, khoffpauir, nmarkle {@highpoint.edu}

*Abstract* — Honeypots are a deceptive technology used to capture malicious activity. The technology is useful for studying attacker behavior, tools, and techniques but can be difficult to implement and maintain. Historically, a lack of measures of effectiveness prevented researchers from assessing honeypot implementations. The consequence being ineffective implementations leading to poor performance, flawed imitation of legitimate services, and premature discovery by attackers. Previously, we developed a taxonomy for measures of effectiveness in dynamic honeypot implementations. The measures quantify a dynamic honeypot's effectiveness in fingerprinting its environment, capturing valid data from adversaries, deceiving adversaries, and intelligently monitoring itself and its surroundings. As a step towards developing automated effectiveness testing, this work introduces a tool for priming a target honeypot for evaluation. We outline the design of the tool and provide results in the form of quantitative calibration data.

*Index Terms* — honeypot, deception, measures of effectiveness, usage, automation

## I. INTRODUCTION

Honeypots are the singular embodiment of deceptive cybersecurity technology. The design is simple: lure would-be-attackers into the system by offering open and vulnerable services. In doing so, practitioners seek to prevent attacks on legitimate systems. Concurrently, researchers capture malicious activity to study with the aim of (a) understanding adversary technologies or (b) developing new countermeasures. Unfortunately, honeypots are difficult to implement correctly [1][2]. Dynamic honeypots are especially prone to configuration issues [3].

To that end, the literature contains a plethora of suggestions as to how honeypots can be effectively deployed or managed. However, there is little quantitative validation of effectiveness in this regard which leaves professionals, researchers, and educators without the means to evaluate implementations or management modalities [4]. While limited work [4] has begun on developing measures of effectiveness, there are still a variety of efforts necessary to realize a usable technology.

One such effort is the technology necessary as a prelude to measure honeypot effectiveness is a mechanism to generate realistic interactions with the honeypot. Prevailing studies have simply exposed honeypots to the public internet or executed ad hoc, pre-canned attacks against a honeypot [5][6][7][8]. These methods are problematic because they are uncontrolled, unpredictable, open to confounding factors, and lack anchors for repeatability study. Accordingly, this work introduces a tool, Primer, which provides controlled, predictable usage generation for honeypot research.

The rest of this work is organized as follows: section two summarize relevant background literature so as to place this work in the proper context; section three describes how we approached the development and calibration of the tool; section four demonstrates the calibration testing results; and, finally, section five documents our conclusions and recommendations.

## II. BACKGROUND

Situating this work in an appropriate context requires a summary of existing honeypot, measures of effectiveness, and TCP/IP replay literature. While a significant amount of literature has been published on honeypots, the interests of the research community have not traditionally included usage generation tools. Accordingly, we provide these discussions with the goal of contextualizing the rationale associated with developing a tool to generate realistic interactions with honeypots. Accordingly, the following sections summarize existing research we found substantive towards our goal.

### A. Honeypots

Honeypots are designed to be attacked, using well known services (e.g. SSH) to attract adversaries [9]. As strictly information gathering technologies, honeypots require careful implementation if the services are going to invite enough adversary activity to produce meaningful data. This is easier said than done [1][2][3].

Honeypots are qualitatively categorized with traits such as interaction level and deployment modes which categorizes them into four types: honeypots, honeynets, honeyfarms, and honeytokens [10]. As well, the initial three can be further categorized according to the degree to which each engages attackers. Lastly, honeypots are largely passive systems with the sole active mechanisms related to adaptation based on surrounding environmental factors [10] or restriction of outbound traffic for legal purposes [11].

Overall, the value of a honeypot rests in the ability to conceal it as a deception technology. Thus, recent research includes suggestions intended to increase the illusory nature of honeypot deployments through five methods [11][12][13]. Regardless, a reliable tool meant to interact with deception technology ought to be loosely coupled to its targets. In doing so, the tool can operate identically and with assurance against a variety of honeypots.

### B. Measures of effectiveness

Measures of effectiveness (MoE) are a relation between performance and specific user requirements [14]. Because MoE must accurately assess whether a tool or service accomplishes a stated task. In the context of this study, such an ability to quantitatively assess and compare honeypots with our MoE [4][15] is a critical motivation for developing a software tool to interact with honeypots so that features can be assessed as MoE. Thus, while the tool resulting from this work does not directly measure effectiveness, it does seek to enable such function.

### C. Network Traffic Replay

The concept and practical behind replaying network traffic are rooted in two use cases. Overall, the concept is to send previously captured traffic to a target endpoint. Further, the type or content of the captured traffic must match the feature or measure intended to be evaluated in the remote endpoint. The evaluation criteria then drive the use cases.

Accordingly, the first use case involves testing networks for diagnostic or troubleshooting purposes [16][17]. This approximates our software but only in a general sense. Accordingly, existing traffic generation tools do not generate realistic interactions with honeypots because the traffic sent to the endpoint is not designed to engage services. Rather, the traffic is designed to elicit error conditions, induce faults, or simply accumulate overhead or computational load [16][17]. In contrast, our goal is to use the network as a means to an end. This use case-diagnostics- places the network as the end.

The second use case also approximates our software but in quite narrow fashion. In this case, the replaying of captured network traffic is used as an attack vector [18][19]. Our intent, while using identical traffic in practical, is not to attack but only generate interactions. Furthermore, we desire return traffic from the target endpoint which precludes tools that do not rewrite checksum fields and so forth.

## III. METHODOLOGY

This work consisted of two phases: development and experimental calibration. The development phase consisted of three subphases and resulted in the Primer software. The experimental calibration followed a two-step protocol with the intended goal of demonstrating robustness as instrumentation for measuring honeypot effectiveness in the future. Data collection and analysis were handled as part of the calibration protocol.

### A. Development

The first step during the development of Primer involved identifying the optimal technology stack. Ultimately, we selected Python as the language because we wanted cross-platform execution with minimal environmental setup.

Then, we identified necessary componentry related to reliably generating network traffic. Here, we gave high priority to repeatability and therefore wanted components which facilitated identical engagements between endpoint targets. Along such lines, replaying pre-canned traffic was a rational decision given the desire for stable, repeatable testing. Thus, we generated a set of packet capture files consisting of ICMP, SSH, and Telnet based attacks.

Subsequently, we developed a module to read in packet capture files, rewrite source and destination IP Address, reset packet checksums, and send the traffic to an indicated endpoint using the scapy library [20]. While there are existing software packages such as TCPReplay [21] which might function in this regard, we found integration and performance to be problematic. As well, most tools do not rewrite the packet header fields necessary to generate realistic traffic towards the target endpoint.

### B. Experimental Calibration Protocol

The first step in the experimental calibration protocol involved generating primary data between Primer and a Cowrie honeypot. The second step involved comparing that data against activity from an Internet-facing Cowrie honeypot.

To facilitate collection of the primary data, we developed an experimental environment consisting of two virtual machines interconnected on a private (host-only) virtual network. The systems ran Ubuntu 18.04 LTS Server with identical base configurations. One system hosted Primer while the second system hosted a Cowrie honeypot implementation with SSH and Telnet services exposed. As well, we deployed a second Cowrie honeypot (also running SSH and Telnet) to an active Internet perimeter.

Analysis of the data consisted of examining the accuracy of network traffic targeting the two honeypots and also the volume of network traffic. We took accuracy to be a manifestation of use or attacks against the honeypots. Volume we conceived as the level of extent of interaction between attacker and honeypot target. Together, these characteristics constitute a validation measure.

## IV. RESULTS

The first set of data analyzed were from the Cowrie honeypot exposed to the Internet. The traffic was captured on the outside interface of the perimeter firewall prior to filtering and NAT. The capture lasted for 30 minutes. Two ports- SSH (22/tcp) and Telnet (23/tcp)- were open. The rest being configured to drop ingress packets. A total of three packets were sent to those ports from two different source systems (Fig. 1). Overall, 17 packets were sent from 12 different source systems to an array of 14 destination ports.

| Address A | Port A | Address B | Port B | Packets | Bytes |
|---|---|---|---|---|---|
| 14.192.212.211 | 51018 | 206.195.147.100 | 445 | 3 | 194 |
| 31.124.112.163 | 49990 | 206.195.147.100 | 23 | 1 | 60 |
| 31.168.191.243 | 58564 | 206.195.147.100 | 81 | 1 | 60 |
| 45.67.14.21 | 35966 | 206.195.147.100 | 22 | 2 | 114 |
| 45.129.33.60 | 42272 | 206.195.147.100 | 30690 | 1 | 60 |
| 45.129.33.60 | 42272 | 206.195.147.100 | 16390 | 1 | 60 |
| 45.129.33.122 | 41118 | 206.195.147.100 | 5957 | 1 | 60 |
| 45.145.66.90 | 49652 | 206.195.147.100 | 2223 | 1 | 60 |
| 45.146.164.169 | 59843 | 206.195.147.100 | 3393 | 1 | 60 |
| 45.146.164.169 | 59843 | 206.195.147.100 | 75 | 1 | 60 |
| 45.146.165.250 | 59757 | 206.195.147.100 | 5097 | 1 | 60 |
| 46.161.27.48 | 43277 | 206.195.147.100 | 23389 | 1 | 60 |
| 51.161.12.231 | 32767 | 206.195.147.100 | 8545 | 1 | 60 |
| 59.126.89.160 | 30579 | 206.195.147.100 | 8080 | 1 | 60 |

Fig. 1: Accuracy of Internet versus Cowrie

The second set of data result from running Primer with two different recorded attacks (executed sequentially) against a Cowrie honeypot. The attacks were issues on an isolated LAN segment to remove confounding factors and general network noise. In both cases, the Primer host shows as 192.168.1.5 while the Cowrie target shows as the 192.168.1.7 host.

The first attack, a Telnet key id overflow, generated 18 packets in less than a second (Fig. 2). The accuracy was higher compared to the general Internet trial but still included extraneous packets such as three sent to port 12235. Such extra traffic is directly attributable to the quality of the source packet capture which Primer replayed. Furthermore, 38% of the packets were replies from the target back to Primer. Such interactivity at a network layer is in direct contrast to the unidirectional traffic seen in the case of the honeypot exposed to Internet traffic.

| Address A | Port A | Address B | Port B | Packets | Bytes |
|---|---|---|---|---|---|
| 192.168.1.5 | 52234 | 192.168.1.7 | 22 | 2 | 108 |
| 192.168.1.5 | 41085 | 192.168.1.7 | 23 | 6 | 360 |
| 192.168.1.5 | 43691 | 192.168.1.7 | 12235 | 3 | 222 |
| 192.168.1.7 | 52234 | 192.168.1.5 | 22 | 2 | 120 |
| 192.168.1.7 | 41085 | 192.168.1.5 | 23 | 5 | 301 |

Fig. 2: Primer accuracy with Telnet key_id attack.

The second attack focused on SSH. Given that Cowrie is primarily used as a SSH honeypot, there was a sound rationale for scoping a calibration trial specifically to that service. The results (Fig. 3) reveal perfect accuracy of sent and received packets between the host and target.

| Address A | Port A | Address B | Port B | Packets | Bytes |
|---|---|---|---|---|---|
| 192.168.1.5 | 36269 | 192.168.1.7 | 22 | 13 | 2939 |
| 192.168.1.7 | 36269 | 192.168.1.5 | 22 | 13 | 2347 |

Fig. 3: Primer accuracy with SSH attack.

After completing the accuracy analysis, we examined the volume of traffic. The first observation centered on the volume of traffic associated with Internet sources interacting with the perimeter Cowrie honeypot. Overall, we found the volume to be low and sporadic. Not only were few packets sent to the honeypot but the time between packets was extensive- between 30 and 50 seconds regularly.

Comparatively, the volume of network traffic exhibited during the Primer calibration exercise is more aligned with a legitimate attack (Fig. 4). There is high interactivity front-loaded and then an abrupt cessation once the attack is finished. This is visible in the large volume of network traffic sent from Primer to the honeypot over a short time. Specifically, Primer sent 13 packets in two hundredths of a second.

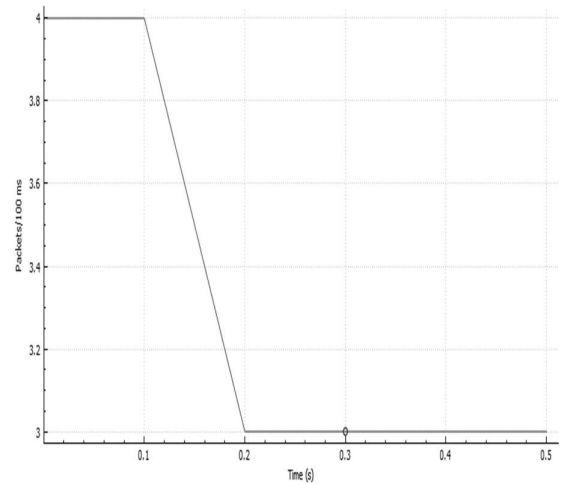

Fig. 4: Primer traffic volume targeting Cowrie

V. CONCLUSION

Given the results, Primer appears to produce accurate, high volume data in short, controlled bursts which targets only honeypot services. This was the intended performance profile. In contrast, exposing a honeypot to the Internet produces erratic, low volume traffic patterns which do not appear to target honeypot services specifically. Thus, there is no guarantee of an attack to elicit interaction and utilization in a honeypot.

This work demonstrates how Primer may eliminate the shortcomings of relying on the Internet to test honeypot implementations. Primer can achieve this by ensuring that the honeypot is attacked in a predictable, consistent manner. As well, Primer allows validation for the MoE of a honeypot as researchers or practitioners will be able to analyze attacks performed on the honeypot and the honeypot's reaction.

Future work involving the use of Primer to drive analysis of specific honeypot MoE may be of benefit. Such study might select a limited set of MoE based on prior taxonomic structure [4] and develop the uniform quantitative measures. Along those lines, future study related to using Primer to test honeypots already deployed in production environments could be of benefit to practitioners. This research might seek to measure potential differences between expected honeypot behavior given implementation characters and actual honeypot behavior.